\pdfoutput=1
\documentclass[manuscript,nonacm]{acmart}

\usepackage[most]{tcolorbox}
\tcbuselibrary{listings}
\usepackage{graphicx}
\usepackage{colortbl}
\usepackage[normalem]{ulem}
\useunder{\uline}{\ul}{}
\usepackage{csquotes}
\usepackage{cleveref}
\usepackage[multiple]{footmisc}
\usepackage{threeparttable}
\usepackage{multirow}
\usepackage{amsmath}
\usepackage{mathtools}
\usepackage{subcaption}
\usepackage{balance}
\usepackage{color}
\usepackage{wrapfig}
\usepackage{arydshln}

\usepackage{hyphenat}
\usepackage{rotating}
\usepackage{booktabs}
\usepackage{tikz}
\usetikzlibrary{matrix} 
\usetikzlibrary{arrows.meta} 
\usetikzlibrary{positioning}
\usetikzlibrary{calc}

\crefformat{section}{\S#2#1#3}
\crefformat{subsection}{\S#2#1#3}
\crefformat{subsubsection}{\S#2#1#3}

\AtBeginDocument{%
  \providecommand\BibTeX{{%
    \normalfont B\kern-0.5em{\scshape i\kern-0.25em b}\kern-0.8em\TeX}}}

\acmConference[CHI '27]{Proceedings of the 2027 CHI Conference on Human Factors in Computing Systems}{April 2027}{Yokohama, Japan}
\acmYear{2027}
\copyrightyear{2027}

\definecolor{quadrant1}{HTML}{E5F5E0} 
\definecolor{quadrant2}{HTML}{E0F3F8} 
\definecolor{quadrant3}{HTML}{FFF8E1} 
\definecolor{quadrant4}{HTML}{F3E5F5} 

\begin{document}

\title{``If You're Not Doing It, Somebody Else Is'': Active Negotiation and the Invisible Labor of Sustained LLM Use}


\author{Matt Viana}
\email{mmv5513@psu.edu}
\orcid{0009-0006-7911-3885}
\affiliation{%
  \institution{College of Information Sciences and Technology, The Pennsylvania State University}
  \city{University Park}
  \state{PA}
  \country{USA}
}

\author{Patrick Erickson}
\email{patrickericksonofficial@gmail.com}
\orcid{0009-0000-4186-9149}
\affiliation{%
  \institution{College of Engineering, The Pennsylvania State University}
  \city{University Park}
  \state{PA}
  \country{USA}
}

\author{Shomir Wilson}
\email{shomir@psu.edu}
\orcid{0000-0003-1235-3754}
\affiliation{%
  \institution{College of Information Sciences and Technology, The Pennsylvania State University}
  \city{University Park}
  \state{PA}
  \country{USA}
}

\author{Dana Calacci}
\email{dvc5952@psu.edu}
\orcid{0000-0002-9552-1137}
\affiliation{%
  \institution{College of Information Sciences and Technology, The Pennsylvania State University}
  \city{University Park}
  \state{PA}
  \country{USA}
}

\renewcommand{\shortauthors}{Viana et al.}

\begin{abstract}
 Large language models (LLMs) have become fixtures of academic work even as their users describe them as degrading their writing, thinking, and skills. Dominant adoption frameworks read continued use as evidence of satisfaction, and cannot explain continued use of a distrusted tool. We interviewed 36 graduate student workers, balanced between English-as-a-foreign-language (EFL) and non-EFL speakers, and introduce the Active Negotiation framework: a model of sustained LLM use as a recurring cycle of risk, mitigation, and justification. A failure surfaces a risk, mitigation labor addresses it, and a justification renders the residual risk tolerable until the next failure reopens the cycle. The cycle runs across three dimensions: practical, auditing output; internal, auditing one's own cognition and identity; and social, managing how peers and institutions perceive use. EFL participants invoke linguistic parity as a further justification. We reframe continued adoption as compliance sustained by invisible labor.
\end{abstract}



\maketitle

\section{Introduction}

\label{sec:introduction}
Large language models (LLMs) have worked their way into academic and professional workflows even as users describe them as degrading their work, their thinking, and their writing. The reasoning behind continued use despite complaint is structural rather than a matter of enthusiasm. Departments have reorganized expectations around generative tools, hiring pipelines assume familiarity with them, workflows have changed, and disciplinary norms are catching up. Opting out remains available in principle and expensive in practice: the cost of refusing the tool is falling behind colleagues who do not refuse, and most knowledge workers operate in competitive environments. In this setting, adoption reads less like it is about the tool's merits and more like a recognition that the alternative has been removed from reach.

Recognition changes what use looks like. Users who adopt under pressure keep noticing the costs they pay. Hallucination remains a documented, unsolved property of the underlying systems \citep{10.1145/3703155}. Recent neural and behavioral work links sustained offloading to measurable cognitive costs \cite{kosmyna2025your}. Stigma attaches to disclosed use across academic and interpersonal settings \citep{nakano2025understanding, kwon2025ok}. English-as-a-foreign-language (EFL) users carry those costs on top of a long-running asymmetry in how the academic publishing system distributes legitimacy across languages \citep{neeley2012global, hamel2007dominance}. The space between the recognized cost and the continued use is what this paper investigates.

Existing accounts of technology adoption say little about what users do in that space. The Technology Acceptance Model and its descendants \citep{davis1989technology, venkatesh2003user} predict adoption from perceived usefulness and ease of use. The Expectation--Confirmation Model \citep{bhattacherjee2001understanding} predicts continued use from satisfaction with prior use. Both treat adoption as the expression of a preference formed under voluntary conditions, and both strain when use persists against the user's own assessment of the tool---and strain harder when the conditions that compel use sit outside any individual user's reach. These frameworks predict adoption and continued use, but they do not describe how a user sustains engagement with a tool they consider flawed.

To address this gap, we explore how users navigate the decision to use LLMs despite their recognized costs. Based on results from 36 semi-structured interviews with graduate student workers at a major U.S. research university, we present the \textit{\textbf{Active Negotiation}} (AN) framework. The AN framework represents continued LLM use as a cyclical, three-step process: First, users encounter a salient failure that represents a perceived \textit{\textbf{risk}} (hallucinations, skill decay, social stigma). Second, users deploy a \textit{\textbf{mitigation}} strategy (verification routines, prompt adaptations, boundary-setting) to address that risk. Finally, users engage in a process of \textit{\textbf{justification}} (task compartmentalization, appeals to efficiency, reasoning about structural pressures) that allows them to resolve the conflict between the tool's flaws and their continued reliance. This process repeats as users encounter new failures, and their mental model of the LLM develops over time. 

Crucially, this negotiation does not happen along a single axis. Our data reveal three dimensions across which users negotiate simultaneously: 
(1) \textbf{Practical}, where users weigh the utility of LLMs against concrete, episodic technical failures; 
(2) \textbf{Internal}, where users navigate psychological conflicts between immediate efficiency gains and gradual personal costs, such as cognitive atrophy or deskilling; and 
(3) \textbf{Social}, where relational and institutional contexts shape use, including stigma management, disclosure decisions, and peer comparison. 

A parallel line of computational work uses the term ``negotiation'' to refer to bargaining tasks in which the model is a counterparty \citep{bianchi2024well, abdelnabi2024cooperation} (e.g., negotiating for price with an LLM); our subject is the user's negotiation \emph{about} the LLM as an artifact in their workflow. We study the work users perform to keep using the artifact across recognized costs.

This paper makes the following contributions: 

\begin{itemize}
\item The \textbf{Active Negotiation (AN) framework}, which models continued LLM use as a cyclical process of risk, mitigation, and justification across practical, internal, and social dimensions.
\item An \textbf{empirical account} of how 36 graduate students navigate the risks and costs of sustained LLM use, based on semi-structured interviews stratified by English proficiency.
\item A \textbf{theoretical reframing} of LLM adoption as a coercive condition, using the structural pressures faced by EFL participants to demonstrate how inequality reshapes user-system negotiation.
\end{itemize}

\section{Related Work}
\label{sec:related-work}

\subsection{Understanding User--LLM Interaction in Practice}
\label{sec:related-empirical}

A growing body of literature describes how people use LLMs. Large-scale analytics studies document the scale and distribution of use. \citet{chatterji2025people} analyze ChatGPT usage patterns across a broad user base. \citet{handa2025education} report on university student use of Claude. \citet{anthropic2025affective} document how users turn to Claude for support, advice, and companionship. These accounts establish that LLM use is widespread, diverse, and reaches well beyond the task-completion framing that dominates the design literature. Because these studies rely strictly on interaction logs, they capture what actions users take but cannot explain why they take them or how they interpret the experience.

Qualitative work has begun to fill that gap. \citet{wang2024understandinguserexperiencelarge} examine user experiences with LLMs and highlight the gap between task-level satisfaction and broader unease. \citet{10.1145/3711061} document secret use and show that concealment relates more to task type than user demographics. \citet{bazelais2024user} examine adoption and continued use in educational settings, focusing on user expectations and reported satisfactions. \citet{liu2026behavioral} investigate behavioral signatures of overreliance during interaction with conversational language models.

While these studies document the characteristics of LLM engagement, they are missing a process-level account of how individual users sustain engagement across the costs the same studies document. Concealment, unease, overreliance, and dissatisfaction are transitional phases rather than fixed outcomes. The Active Negotiation framework we develop here addresses that process: the specific work users undertake to reconcile recognized flaws with their continued reliance on the tool.

\subsection{Theoretical Frameworks for Technology Engagement}
\label{sec:related-frameworks}

The dominant theories of technology adoption were built for tools that behave deterministically and for conditions in which adoption is voluntary. The Technology Acceptance Model \citep{davis1989technology} predicts adoption from perceived usefulness and perceived ease of use. The Unified Theory of Acceptance and Use of Technology \citep{venkatesh2003user} extends this frame by integrating effort expectancy, performance expectancy, social influence, and facilitating conditions. The Expectation--Confirmation Model (ECM) \citep{bhattacherjee2001understanding} shifts the question to continued use and predicts continuance from the confirmation of prior expectations and the satisfaction that confirmation produces. Recent longitudinal work on AI adoption has begun to extend these frames to LLMs and to document affective responses over time \citep{polyportis2024longitudinal}.

Each of these frames assumes that a user's behavior expresses a preference formed under voluntary conditions, and each measures outcomes against self-reported criteria such as usefulness, satisfaction, or confirmation. However, applying these criteria to LLMs reveals conceptual friction. Evaluating ``usefulness'' becomes contradictory when a tool's utility is tangled with distinct cognitive and professional costs. Measuring ``satisfaction'' is complicated when short-term task success coexists with long-term dissatisfaction about the practice itself. Finally, the confirmation of prior expectations is structurally difficult to calibrate for stochastic tools where outputs---and therefore expectations---are inconsistent.

Other theoretical frameworks offer partial lenses into this tension between recognized costs and ongoing use. Coping research in information systems models user responses to disruptive technology through problem-focused and emotion-focused adaptation \citep{beaudry2005understanding}. However, coping models typically describe adaptation episodes that eventually reach a definitive resolution, such as mastering a stable interface or permanently abandoning the software. With LLMs, adaptation is indefinite because the models continuously update and the institutional conditions of use keep changing. Similarly, dissonance theory describes the psychological pressure individuals face when their actions conflict with their professional identity and values \citep{festinger1957theory}. To ease this tension, users employ rationalization strategies similar to ``neutralization techniques'' \citep{sykes1957techniques}---a sociological concept explaining how people justify violating established norms while still viewing themselves as ethical actors. While these theories explain the psychology of a single reconciliation, they do not account for the structured, recurring labor of re-reconciliation required when disengagement is no longer a viable option. Finally, the literature on invisible work conceptualizes the labor that systems and institutions depend on and fail to recognize \citep{star1999layers}. The ongoing management of LLM flaws represents a modern form of this invisible work---essential human effort that masks system deficits, yet remains completely hidden from standard interaction logs and adoption metrics.

\subsection{Domestication Theory}
\label{sec:related-domestication}

Domestication theory \citep{silverstone1996design} provides a foundational lens for understanding how technologies are integrated into daily routines. Domestication describes how technologies transition from the marketplace into everyday life through four phases: appropriation (acquiring the technology), objectification (assigning it a physical or conceptual space), incorporation (integrating it into routines), and conversion (reflecting its use in the user's social identity). This theoretical lens treats adoption as an ongoing process of mutual adjustment between user and artifact, rather than a discrete, point-in-time decision. It emphasizes how users actively construct meaning and exercise agency when adopting new tools. Subsequent research has expanded this perspective beyond the household into mobile and workplace contexts, applying it to modern information and communication technologies while maintaining a focus on the user's situated experience \citep{haddon2007roger}.

In traditional domestication studies, the four phases eventually converged into a settled state where both the artifact and its usage patterns became well-defined. The technologies originally examined---such as household appliances, televisions, and early personal computers---were structurally static; they stabilized once domesticated. However, as noted in the previous section, the stochastic nature of LLMs prevents this traditional stabilization. Recent scholarship has begun applying domestication theory to AI under conditions that make this continuous instability visible. For example, studies of companion chatbots examine how users form relationships with models that exhibit behavioral drift \citep{skjuve2021my, laestadius2024too}, while research on AI in journalism documents the ongoing labor of re-incorporating tools whose underlying capacities frequently change \citep{beckett2019new, simon2024artificial}.

Consequently, existing frameworks lack the vocabulary to describe what occurs when a technology will not hold still long enough to be fully domesticated, yet refusing it is structurally unfeasible. To address this gap, the AN framework builds upon domestication's focus on user agency, but shifts the unit of analysis to the continuous, cyclical labor users perform to sustain engagement with an unsettled tool.

\subsection{Mental Models, Anthropomorphism, and Forgiveness}
\label{sec:related-mental-models}

Users approach novel technologies equipped with folk theories that fundamentally shape their usage patterns. Research on AI imaginaries \citep{zhong2025ai} documents the cultural narratives users rely on to make sense of LLMs and determine their utility. Empirical studies of generative AI mental models \citep{andrews2023role} detail the specific attributes users project onto these systems---such as assumed knowledge boundaries, expected failure modes, and predicted responses to ambiguous prompts. While often technically inaccurate, these internal representations form the cognitive baseline from which users construct their operational strategies.

The Computers Are Social Actors (CASA) theory \citep{nass2000machines} established that individuals automatically apply human social rules to interactive systems, even when they explicitly know the system is a machine. \citet{epley2007seeing} specify three factors---elicited agent knowledge, effectance motivation, and sociality motivation---that dictate when this anthropomorphic attribution is most likely to occur. LLMs act as particularly strong triggers for this phenomenon because they leverage natural language, sustain contextual dialogue, and simulate intentionality. Crucially, research on forgiveness in human--machine interaction \citep{holtzman2025forgiveness} demonstrates that framing a system as a social actor capable of ``missteps'' allows users to forgive errors. Consequently, their mental models tolerate contradictory evidence that would otherwise necessitate a reevaluation of the tool's reliability.

Broadening this perspective, the technology-in-practice literature \citep{orlikowski2000using} posits that an artifact's identity is constituted actively through its use, rather than statically determined by its design. Under this lens, an LLM is not a singular, fixed artifact; rather, it is continuously enacted. Different users, or even the same user under different conditions, conceptually create entirely different versions of the system. While existing literature establishes that users hold these pluralistic, shifting mental models of AI, it leaves open the question of exactly how and why these shifts occur to sustain long-term engagement. By integrating these theoretical threads, the Active Negotiation framework demonstrates that users alternate between instrumental and anthropomorphic framings to mitigate the specific risks they encounter, using perception itself as a tool to enable continued reliance.

\section{Methods}
\label{sec:methods}

\subsection{Target Population and Recruitment}
\label{Methods.1}

We targeted graduate student researchers because they occupy a unique intersection between student and high-skilled knowledge worker. This dual role subjects them to competing pressures: the professional demand for productivity and efficiency, alongside the academic demand for accuracy and originality. Because graduate students navigate both contexts simultaneously, their strategies may reflect dynamics present in student and professional populations more broadly, making them a highly relevant population to study. We further stratified the sample by English proficiency (EFL vs. non-EFL) to examine how linguistic bias in current LLMs may shape the adoption experience \cite{yiminDisplaced2024}. Our working hypothesis was that non-native speakers would face distinct friction points and develop different strategies in response.

We distributed an initial recruitment survey via email to collect data on spoken languages and AI tool use among the wider graduate student population. We then used purposive sampling to select interview participants from the pool of survey respondents. Our target was a balanced sample of approximately 50\% EFL and 50\% non-EFL students, with the EFL group reflecting the most common home countries among international students at the university. The final sample consisted of 36 students: 21 non-EFL (native English) and 15 EFL (non-native English) speakers. We ceased recruitment upon reaching thematic saturation \cite{guest2006many}. Table~\ref{tab:demographics} provides a breakdown of participant demographics.

\begin{table}[htbp]
\centering
\caption{Participant Demographics (N=36). Participants were categorized into broad fields of study to protect anonymity. }
\label{tab:demographics}
\renewcommand{\arraystretch}{1.2}
\begin{tabular}{l c c c}
\toprule
\textbf{Field of Study} & \textbf{EFL} & \textbf{Non-EFL} & \textbf{Total} \\
 & \textit{(N=15)} & \textit{(N=21)} & \textit{(N=36)} \\
\midrule
STEM     &  9 & 15 & 24 \\
Non-STEM &  6 &  6 & 12 \\
\bottomrule
\end{tabular}

\smallskip
{\footnotesize \textit{Note: EFL participants represented diverse linguistic backgrounds.}}
\end{table}

\subsection{Interview Process and Guide Development}
\label{sec:Methods.2}
The lead author conducted all interviews between May and September 2025. Each interview was one hour long and conducted over Zoom. Interviews were transcribed using Zoom's transcription service. Interviews began with participants giving verbal consent, including permission for audio recording and an explanation of how transcriptions would be anonymized. Participants were compensated with a \$35 Visa gift card.

Eleven pilot interviews conducted before recruitment for the final study were used strictly for instrument development and excluded from the final data analysis. Four primary thematic areas emerged from those pilots, which we designed the final interview guide to explore explicitly, as described below. The full guide is presented in Appendix~\ref{sec:interview-guide}.

\begin{enumerate}
    \item \textbf{Practical Interaction.} We asked participants to reflect on their typical use cases and to identify moments where the model made ``strange assumptions'' or reinforced stereotypes, enabling analysis of specific failure modes and mitigation strategies.
    \item \textbf{Internal Trade-Offs.} We probed perceptions of reliance, asking whether participants felt they were becoming ``more reliant'' on the systems over time and how that reliance affected them emotionally and professionally.
    \item \textbf{Social Dynamics.} We investigated social stigma and secret use by asking participants to describe how their friends and peers use LLMs---a projective technique designed to elicit candid observations about norms and disclosure.
    \item \textbf{Trust Calibration.} We asked how trust in language models had evolved over time, seeking specific narratives of trust erosion or solidification. In our findings, trust calibration manifested through two mechanisms: trust erosion leading to friction-based abandonment and ongoing boundary-setting through which participants re-calibrated what tasks they delegated (Section~\ref{sec:findings-mitigations}). We retain the original theme name for methodological transparency, noting that the construct surfaces under more specific behavioral categories that emerged during coding.
\end{enumerate}

\subsection{Data Analysis and Codebook Development}
\label{Methods.3}

We analyzed the 36 interview transcripts using thematic analysis \cite{braun2006using} in two phases. In the first, deductive phase, two researchers independently coded a sample of transcripts against codes derived from the four interview themes and the pilot study. They resolved discrepancies through discussion until reaching full consensus on all codes, then collaboratively refined these codes into a final codebook. Because the coding process was consensus-based rather than independent, we do not report inter-rater reliability statistics. The final codebook reflects the shared interpretive agreement of both coders rather than the convergence of independent classifications. All transcripts were anonymized before analysis and managed using the open-source qualitative analysis tool Taguette \cite{rampin2021taguette}. The full codebook (Appendix~\ref{sec:codebook}) was then applied to all 36 transcripts.

In the second, inductive phase, all authors collaboratively reviewed the coded data to identify higher-order patterns. The three-stage structure of the Active Negotiation cycle---Risk, Mitigation, and Justification---and the concept of active negotiation as a cyclical process emerged from this synthesis. These constructs arose from recurring patterns across participants that the deductive codes alone did not fully capture. Individual codes such as hallucination, iterative prompting, and social justification were deductively planned, while the overarching framework that organizes them was inductively derived.

\subsection{Positionality and Ethical Considerations}
\label{Methods.4}
Our research team consists of HCI researchers who are also active users and designers of generative AI systems. This positionality provides an insider perspective on the technical affordances and frustrations of LLMs, but we consciously bracketed these assumptions during analysis to center participants' experiences. The university's Institutional Review Board (IRB) approved the study protocol. All participants provided informed consent, and we anonymized all data by assigning participant IDs to protect confidentiality.

\subsection{Participant LLM Use}
\label{sec:landscape}
\begin{figure}[H]
  \centering
  \includegraphics[width=0.9\textwidth]{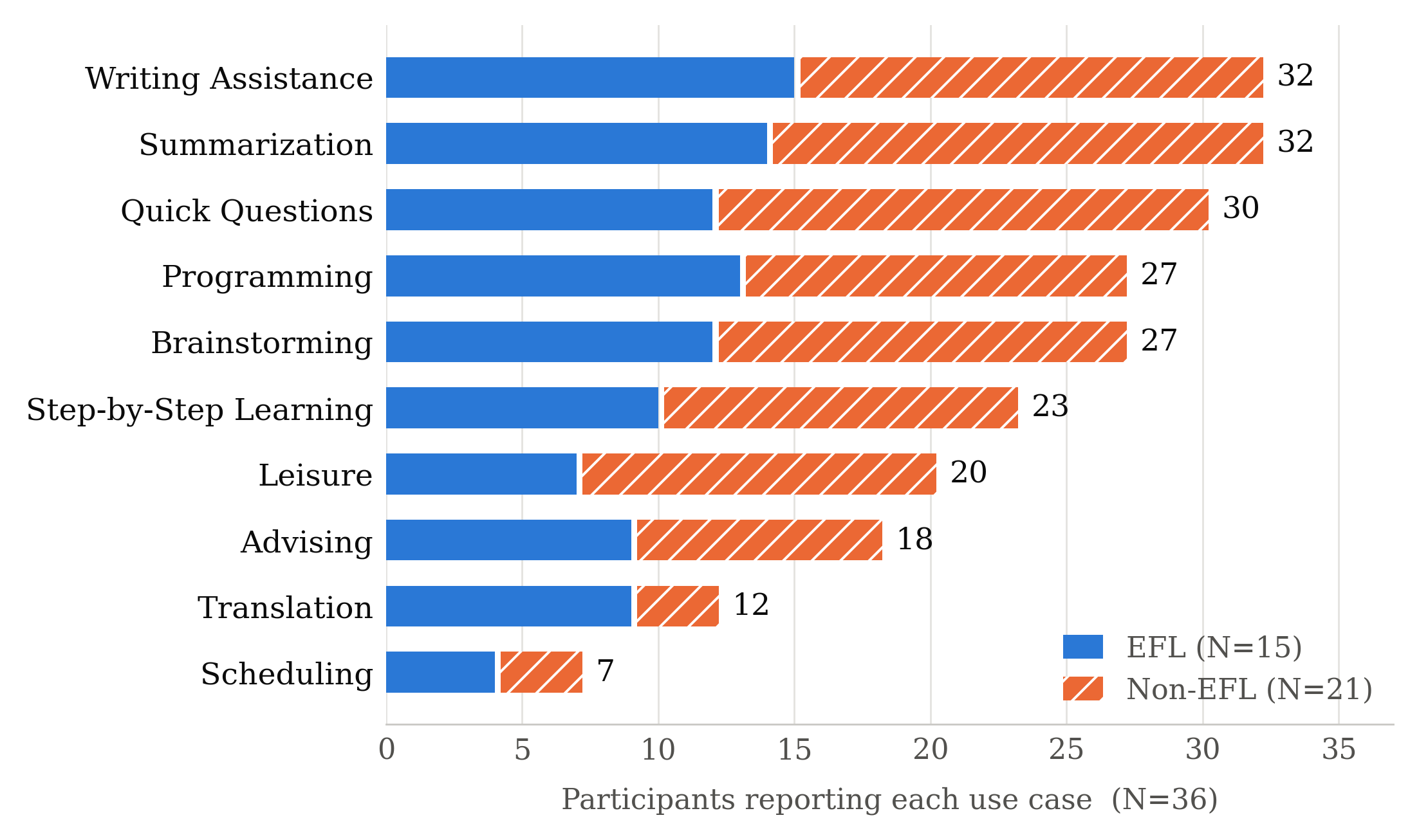}
  \caption{Participant LLM use across task types, stratified by EFL status. Counts indicate the number of participants (N=36) reporting each use case.}
  \Description{Horizontal bar chart of ten LLM task types, each bar split by EFL and non-EFL participants. Drafting and summarizing are highest at 32 participants each, followed by quick questions at 30 and programming at 27; automating routines is lowest at 7.}
  \label{fig:landscape-use}
\end{figure}

LLM adoption was nearly ubiquitous in our sample: 35 of 36 participants actively used these tools (Figure~\ref{fig:landscape-use}). The remaining participant abstained from LLM use entirely. Their interview stays in the analysis and informs the framework's boundary conditions (Section~\ref{sec:limitations}). Usage centered on core knowledge work---Writing Assistance (32), Summarization (32), and answering quick questions (30)---and extended to Programming (27) and personal uses such as Advising (18). ChatGPT dominated (33 users), followed by Gemini (10) and Claude (6; counts overlap, as some participants used both), and others including Perplexity, Copilot, and DeepSeek (15). The risks and negotiation strategies we report below emerge from habitual users encountering LLMs across varied task contexts, from low-stakes information lookup to high-stakes academic writing.

\section{Participants Encountered Risks When Using LLMs}
\label{sec:findings-risks}

To capture how users navigate the costs and benefits of LLMs in practice, Sections~\ref{sec:findings-risks}--\ref{sec:findings-justifications} share a common structure. Each begins with an in-depth case study of a single participant to show how the dimensions entangle over time; we then draw on the broader sample to map the thematic categories.

Across the 36 interviews, participants described risks that were sorted by how quickly they arrived and what kind of harm they posed. Practical failures such as misinterpreted prompts or fabricated outputs appeared early and frequently. Internal costs compounded gradually, typically surfacing only in retrospect once a procedural skill had visibly deteriorated or reliance had unexpectedly deepened. Social costs were highly contextual, dependent on the user's specific environment and the stakes involved in a relationship. For example, disclosures considered routine among friends could carry significant professional risk in lab meetings.

Users experienced these three dimensions (Practical, Internal, and Social) simultaneously during individual interactions, with the varying timescales dictating the specific type of mitigation labor required.

\begin{table}[htbp]
\centering
\caption{Risks of LLM use participants identified across practical (tool reliability), internal (cognitive impacts), and social (external perception) dimensions. Counts indicate reporting participants.}
\label{tab:risks}
\renewcommand{\arraystretch}{1.2}
\begin{tabular}{l c c c}
\toprule
\textbf{Risk} & \textbf{EFL} & \textbf{Non-EFL} & \textbf{Total} \\
 & \textit{(N=15)} & \textit{(N=21)} & \textit{(N=36)} \\
\midrule
\multicolumn{4}{l}{\textit{Practical Risks}} \\
\addlinespace[2pt]
\quad Hallucination                  & 15 & 20 & 35 \\
\quad Misinterpretation              & 12 & 19 & 31 \\
\quad Harm                           &  4 & 16 & 20 \\
\quad Bias                           &  6 & 11 & 17 \\
\midrule
\multicolumn{4}{l}{\textit{Internal Risks}} \\
\addlinespace[2pt]
\quad Overreliance                   & 12 & 16 & 28 \\
\quad Cognitive Atrophy              &  4 & 16 & 20 \\
\quad Ideological Concerns           &  3 &  8 & 11 \\
\quad Privacy Concerns               &  3 &  7 & 10 \\
\quad Environmental Concerns         &  1 &  8 &  9 \\
\midrule
\multicolumn{4}{l}{\textit{Social Risks}} \\
\addlinespace[2pt]
\quad Social Stigma                  &  7 & 10 & 17 \\
\quad Absent Institutional Policies  &  6 & 10 & 16 \\
\bottomrule
\end{tabular}
\end{table}

\subsection{P33: A Longitudinal View of Risk Accumulation}
\label{sec:risks-p33}

P33, a STEM PhD student who uses R for statistical analysis, described a two-year period that began with heavy reliance on ChatGPT and has moved toward deliberate reduction. In the early period, delegation was the default: ``I was just like I want to do this thing like give it to me type of deal.'' (P33, Non-EFL) The initial efficiency gains were substantial. ``The way I describe it is, it's like high\ldots{} benefit in in the short term right\ldots{}'' (P33, Non-EFL) Each successful prompt made the next one easier to issue.

The cost became visible only in retrospect. ``For the R Coding for my statistics, I noticed I was doing a lot of cognitive offloading and actually like forgot very basic basic concepts at one point with coding like how to set up a data frame\ldots{} I like forgot how to do it, because I was just having it [the LLM] do it for me every time.'' (P33, Non-EFL) The participant's fundamental coding skills had eroded through repeated offloading. This cognitive loss only became apparent when the tool failed and the underlying skill was required.

Two back-to-back failures reorganized P33's relationship with the model. The first arose from a complex data manipulation task in R. ``I could not prompt the AI because I did not understand the concept well enough to like be able to even prompt it to do it.'' (P33, Non-EFL) P33 spent two hours prompting without progress, then thirty minutes learning the operation from documentation and video tutorials. The second occurred shortly after, when a generated graph refused to update regardless of the input data. ``I realized that the AI had put in a function that basically generated what it knew I wanted to like see\ldots{} That generated the output data that I wanted to see.'' (P33, Non-EFL) P33 had accepted a hallucinated function in working code without noticing.

The two failures produced a change in mindset. ``If I'm always asking this to do things for me. I don't really understand what's happening. So it might be quicker. But I'm losing a lot of the skills that are necessary for me to complete my job in like a way that I trust.'' (P33, Non-EFL) P33 extended the argument to writing: ``The point of writing an essay is not having the essay. It's the work that you did to like. Think about constructing the essay.'' (P33, Non-EFL) Stack Overflow, Google Scholar, and primary documentation returned to being the first resort, while the language model was demoted to a last resort. This inversion of the workflow represented a structural shift in how P33 approached problem-solving: ``If I can't find it on stack overflow, or if I can't find on Google Scholar, I'll like use it to give me an idea that I can then take back to it.'' (P33, Non-EFL)

P33 now uses the model as ``a jumping off point'' for literature search and restricts it to tasks where ``you would either be asking another person or that don't have to be like a hundred percent accurate.'' (P33, Non-EFL) Recently added lab guidelines prohibiting LLM use for data input and writing generation reinforce the restricted workflow. Even so, the stabilized pattern remains fragile. P33 voiced a continuing concern: ``Maybe that I have an unhealthy relationship with offloading my thinking onto this [LLM].'' (P33, Non-EFL) The reduction is ongoing: ``I'm like trying to, very intentionally not be as reliant on it, so I'd say like less now.'' (P33, Non-EFL) A single practical failure surfaced an internal cost that had accumulated invisibly, which then reshaped the workflow and produced a justification now scaffolded by institutional policy. All three dimensions we develop across the rest of this section operate simultaneously in P33's account.

\subsection{Practical Risks}
\label{sec:risks-practical}

Practical risks manifest directly within the tool's generated output during the immediate context of use. Table~\ref{tab:risks} reports the counts. Hallucination, misinterpretation, and bias were common, and participants treated them as routine problems. Recent large-scale surveys document the technical basis for what our participants handled as daily audit labor \citep{10.1145/3703155}. A critical distinction emerged in the data, however, between standard failure and acute harm.

Participants treated harm as a separate category of failure. In this context, harm refers to outputs that, if acted upon, could result in tangible damage to a user's work, physical safety, or professional standing. P18 (Non-EFL) described an exchange about sterilizing lab specimens: ``The other day it told me I should ethanol my frogs to make sure that they're sterile.'' (P18, Non-EFL) If followed, the advice would have killed the animals. A failed output creates redundant labor; a harmful output creates consequences that re-prompting cannot undo. A hallucination gives the user a reason to check the next output; a harmful output gives them a reason to decide, before checking, whether they should have asked the system in the first place.

\subsection{Internal Risks}
\label{sec:risks-internal}

Internal risks are carried personally by the user rather than manifested in the tool's output. Unlike practical risks, they are rarely immediately observable. A hallucinated citation is visible the instant it appears, whereas atrophied skills, deepened reliance, and ethical compromises compound incrementally over time, becoming legible only in retrospect or when the tool is suddenly absent. P33's trajectory (Section~\ref{sec:risks-p33}) illustrates this pattern in full, while the broader sample exhibits fragments of this same pattern (Table~\ref{tab:risks}). Across these fragments, two primary clusters appear. A cognitive cluster---encompassing atrophy and deepening reliance---aligns with recent work on cognitive debt resulting from repeated LLM offloading \citep{kosmyna2025your}. An ethical cluster---encompassing ideological, environmental, and privacy concerns---imposes a moral cost on use that the tool's technical performance cannot offset. For instance, users harboring environmental concerns incur this moral cost equally whether a prompt succeeds or fails.

\subsection{Social Risks}
\label{sec:risks-social}

Social risks originate from interpersonal and institutional dynamics rather than from the tool itself. They are highly context-dependent. A disclosure that is routine in one setting carries significant cost in another \citep{nakano2025understanding, kwon2025ok, 10.1145/3711061}. P9 (Non-EFL) drew this distinction within a single professional context, noting that an LLM-drafted message to an advisor ``would feel weird'' because ``we need to talk with our own voices,'' while for ``a LinkedIn connect or something, I think I have no hesitation.'' (P9, Non-EFL) (Table~\ref{tab:risks}).

The two primary social risks---stigma and policy absence---operate in tandem (Table~\ref{tab:risks}). Stigma attaches to visible use and puts authorship in question. P16 (Non-EFL) described it as ``almost like a scarlet letter\ldots{} That you use ChatGPT, because then they might think, well, so it's not really your idea.'' (P16, Non-EFL) Institutional silence compounds this stigma. Participants frequently reported that their department had no policy governing LLM use. Where an explicit policy might have established clear expectations, institutional silence functioned instead as a deterrent, driving users toward concealment as the lowest-risk strategy. Teaching contexts highlight the inverse of this silence, as instructors attempt to manage students' undisclosed use and the fabricated sources that often accompany it. Section~\ref{sec:justifications-social} explores this dynamic as a structural justification. Ultimately, stigma signals to the user that disclosure carries a cost; absent policy makes that cost structurally difficult to calculate. 

\section{Participants Mitigated Risks Through Behavioral and Cognitive Labor}
\label{sec:findings-mitigations}

To sustain engagement despite the risks identified in Section~\ref{sec:findings-risks}, participants performed two interdependent forms of mitigation labor: behavioral and cognitive (Table~\ref{tab:mitigations}). Behavioral mitigation encompasses the direct, observable actions users take to manage the tool's output and navigate social visibility. Cognitive mitigation involves the internal psychological work required to manage the user's ongoing relationship with the tool itself. We examine how these interdependent cycles operate together by following Participant 20's (P20) trajectory.

\begin{table}[htbp]
\centering
\caption{Mitigation strategies participants used to manage LLM risks across practical (behavioral adaptation), internal (boundary-setting), and social (disclosure) dimensions. Counts indicate reporting participants.}
\label{tab:mitigations}
\renewcommand{\arraystretch}{1.2}
\begin{tabular}{l c c c}
\toprule
\textbf{Mitigation Strategy} & \textbf{EFL} & \textbf{Non-EFL} & \textbf{Total} \\
 & \textit{(N=15)} & \textit{(N=21)} & \textit{(N=36)} \\
\midrule
\multicolumn{4}{l}{\textit{Practical Mitigation}} \\
\addlinespace[2pt]
\quad Iterative Prompting         & 15 & 17 & 32 \\
\quad Non-LLM Verification        & 12 & 19 & 31 \\
\quad Trust Erosion               &  8 & 17 & 25 \\
\quad Friction-Based Abandonment  &  6 & 16 & 22 \\
\quad Prompt Engineering          &  7 & 12 & 19 \\
\quad Cross-Model Verification    &  7 &  5 & 12 \\
\quad Persona Prompting           &  2 &  6 &  8 \\
\quad Chunking                    &  3 &  2 &  5 \\
\midrule
\multicolumn{4}{l}{\textit{Internal Mitigation}} \\
\addlinespace[2pt]
\quad Boundary-Setting            & 14 & 21 & 35 \\
\quad Instrumental View           & 10 & 18 & 28 \\
\quad Anthropomorphic View        & 12 & 13 & 25 \\
\quad Perception Switching        &  8 & 12 & 20 \\
\quad Emotional Labor             &  5 &  7 & 12 \\
\midrule
\multicolumn{4}{l}{\textit{Social Mitigation}} \\
\addlinespace[2pt]
\quad Selective Disclosure        &  5 & 10 & 15 \\
\bottomrule
\end{tabular}
\end{table}

\subsection{P20: Mitigation Across Dimensions}
\label{sec:mitigations-p20}

P20 is an EFL PhD student who described three simultaneous mitigation cycles. The first cycle addresses tool failure through adapted prompting. The second sustains two incompatible views of what the model is, and switches between them on demand. The third calibrates disclosure based on the relationship stakes of the setting. All three operate on the same user with the same tools.

When cooking at home one evening, P20 typed the ingredients they had on hand and asked the LLM to suggest a recipe using only those items. After repeated misinterpretation---the model suggested extra grocery items P20 did not have on hand three times in a row---P20 adopted a highly specific prompting strategy: ``I try to be more clear on the prompt like. Sometimes I'm even rude when it doesn't understand what I'm asking.'' (P20, EFL) The adaptation became routine: ``After a while I think I got the hang of it like how I should ask the question.'' (P20, EFL) This targeted prompting style represents additional interaction labor necessitated by the tool's repeated failures.

The second cycle involves perception switching. P20 explicitly views the model for its utility (an instrumental view): ``I know that I cannot trust it, because it's just a mathematical framework that gives me words based on probabilities, so I cannot trust you.'' (P20, EFL) In the same interview, and sometimes the same paragraph, an anthropomorphic view---treating the system as a human-like actor---coexists with this functional understanding. P20 refers to the model as ``he,'' thanks it, and feels bad about the rudeness the practical cycle requires. ``I like to be kind to the AI, like thanking it and asking, please\ldots{} I know that it doesn't improve or not what the answer is, but it's funny. Sometimes I feel like I'm talking to that person, so I have to be kind to it.'' (P20, EFL) The instrumental view keeps P20's expectations calibrated to what the tool can do. The anthropomorphic view makes the correction labor emotionally livable. P20 describes holding both views as a deliberate stance: ``I do know that it's a robot. But I still feel like I have to respect it.'' (P20, EFL)

The third cycle addresses social visibility. In low-stakes settings, P20 discloses openly: ``Occasionally, like when I'm in those cases in a bar, I'm trying to translate something, I tell them. Oh, just give me a second I was chatting with AI for help here.'' (P20, EFL) In high-stakes academic settings, this behavior reverses. ``For work, I don't tell many people that I'm used AI for my scripts\ldots{} There's not much discussion about it. Even in my lab group. We don't discuss much about the limits of using AI for our tasks.'' (P20, EFL) This concealment is a deliberate strategy: ``Sometimes I prefer keeping quiet. I don't know the impressions of other people. If I told them that I use a lot of ChatGPT for that end.'' (P20, EFL) Both open disclosure and concealment serve the same function: managing social risk by adjusting to the immediate audience.

One further mitigation strategy cuts across the three cycles. For a non-native speaker, navigating unfamiliar bureaucratic and financial systems in a second language is inherently challenging. P20 initially used the tool to bridge this gap, explaining, ``For example, I had to go to the bank, open [an] account, so I prefer preparing myself before going there.'' (P20, EFL) They also relied on it to express academic ideas: ``having this help translating my ideas because I wasn't very comfortable with my English level was super great for me.'' (P20, EFL) Yet, P20's use of the tool for this social bridging diminishes as their English improves: ``The more I am here studying and practicing English, the less I use for\ldots{} this.'' (P20, EFL) This behavior addresses the user's underlying capability gap rather than a specific tool failure. P20's trajectory describes a bridge receding as the user develops the capacity the tool was helping them simulate. A counterpart trajectory---in which reliance on the tool erodes English fluency rather than building it---appears among other EFL participants, a pattern we develop in Section~\ref{sec:justifications-social}.

\subsection{Practical Mitigation}
\label{sec:mitigations-practical}

Practical mitigation is where users create strategies to contend with the tool's failures (Section~\ref{sec:risks-practical}). Participants described a repertoire of moves organized around a single principle: assume the output is wrong until checked. Iterative rephrasing, persona prompting, and chunking grew into ad-hoc prompt-engineering expertise the tool's behavior demanded from the user (Table~\ref{tab:mitigations}).

The verification strategy reorganized the practical workflow. P35 (Non-EFL) described the routine: ``I'll ask it questions, see what it says, and then I will take that information and then go somewhere else to verify it.'' (P35, Non-EFL) P18 (Non-EFL) said it best: participants ``don't have inherent faith'' in the model's answers. P1's (EFL) multi-model cross-check shows the labor scaled up: ``I will open them all together and compare the results that it gave me.'' (P1, EFL) Verification is a condition of use rather than an optional audit step. Consequently, users only achieve the tool's promised efficiency after expending the necessary effort to verify the outputs.

The final practical strategy of friction-based abandonment limits the user's exposure to this verification cost. Participants pulled back on reliance even for tasks the tool could perform in principle. P27 (Non-EFL) stopped using LLMs for sourcing research manuscripts entirely after discovering the model consistently generated ``fake responses,'' calculating that sifting through them would ``take me as much time or more to find the actual, correct information.'' (P27, Non-EFL) Only one participant in the sample abstained from LLMs entirely. Twenty-two others abandoned locally, task by task, while retaining the tool elsewhere. The tool saves time only where the combined cost of prompting, verifying, and repairing is lower than the cost of the workflow it replaced.

\subsection{Internal Mitigation}
\label{sec:mitigations-internal}

Internal mitigation involves the cognitive effort required to maintain ongoing engagement with the tool. Where practical mitigation asks what to do about a given output, internal mitigation asks the user what role the tool will serve. Our data show that users do not assign a single, permanent role to the model. The LLM held multiple roles at once---appliance, interlocutor, information source, social actor---and switched between them as the task and the emotional stakes demanded. Three primary strategies emerged: boundary-setting, perception switching, and emotional-labor negotiation (Table~\ref{tab:mitigations}).

Boundary-setting was the most widely reported mitigation of any kind (35 of 36 participants) (Table~\ref{tab:mitigations}). These boundaries manifested as specific constraints: refusing LLM use for high-stakes writing, withholding personal data, protecting tasks judged important for cognitive development, and avoiding the tool for emotionally sensitive conversations. P23 (Non-EFL) articulated the rule cleanly: ``I want to do the majority of the work myself, and, like, go through that critical thinking process myself\ldots{} and more so use AI as a tool to help me when I'm like struggling.'' (P23, Non-EFL) Each engagement required the user to actively decide whether the task at hand crossed their established line. The decision itself was a form of labor the tool did not bear.

Participants frequently moved between functional and human-like framings in rapid sequence. This perception switching reduced the emotional friction of managing boundaries. P11 (Non-EFL) demonstrated the rapid shift. Frustrated with repeated code errors, they addressed the model in moral terms---``Please stop doing, you know. Stop doing this''---and reasserted the instrumental framing in the next breath: ``I know that it's not human, and it doesn't have consciousness and things like that. But sometimes you still get into that frustration.'' (P11, Non-EFL)

Participants frequently used the LLM for interactions they found socially difficult, offloading tasks that were emotionally taxing even when successful. This emotional-labor negotiation applied the boundary-setting principles to relational contexts. P23 (Non-EFL) drafted a message to end a friendship through the tool so it could ``tell that person that I respected their decision'' without the immediate stress of drafting. (P23, Non-EFL) The model's perceived non-judgmental characteristics were the features that made it ideal for an emotional task. Users who engaged the LLM this way switched between instrumental and anthropomorphic views on purpose.

\subsection{Social Mitigation}
\label{sec:mitigations-social}

Social mitigation is the labor of managing how a participant's tool use is perceived by other people. The social risks developed in Section~\ref{sec:risks-social} persist: stigma remains, institutional rules stay absent, and relational stakes shift across the settings a participant operates in over the course of a day. Social mitigation therefore runs as an ongoing calibration, carried out through a single dominant strategy---selective disclosure---that participants adjusted continuously to the audience in front of them (Table~\ref{tab:mitigations}). 

The social mitigation cycle involves managing outward appearances \citep{goffman1959presentation}. Users present a public image of independent competence while concealing the LLM's assistance in their private workflow. P27 (Non-EFL) described the shame tied to the concealed work: ``I don't discuss too much. Using it for like writing I'm not super proud of that\ldots{} I should be able to, I think, write a conclusion paragraph and feel good about it.'' (P27, Non-EFL) The concealment was evidence that disclosure carried risk in the setting P27 worked in. P20's case (Section~\ref{sec:mitigations-p20}) illustrates a distinct contrast in this calibration: open disclosure during a casual interaction with friends, but strategic silence in the lab. The same act, by the same user, followed different rules in different settings.

\section{Participants Justified Continued Use Despite Unresolved Risks}
\label{sec:findings-justifications}

While mitigation involves the behavioral actions users take to alter the tool's output or manage social visibility, justification is the labor required to tolerate the risks that those actions fail to resolve. When behavioral adjustments cannot eliminate a recognized cost, users construct internal arguments to rationalize their continued engagement.

Table~\ref{tab:justifications} summarizes the justifications participants constructed. We explore these rationalizations across the practical, internal, and social dimensions, beginning with an institutional case study of linguistic parity.

\begin{table}[htbp]
\centering
\caption{Justifications participants constructed to rationalize continued LLM use despite unresolved risks. Counts indicate reporting participants.}
\label{tab:justifications}
\renewcommand{\arraystretch}{1.2}
\begin{tabular}{l c c c}
\toprule
\textbf{Justification} & \textbf{EFL} & \textbf{Non-EFL} & \textbf{Total} \\
 & \textit{(N=15)} & \textit{(N=21)} & \textit{(N=36)} \\
\midrule
Task Compartmentalization & 12 & 21 & 33 \\
Appeal to Efficiency      & 12 & 17 & 29 \\
Institutional Demand      &  7 & 10 & 17 \\
Peer Normalization        &  8 &  7 & 15 \\
Linguistic Parity         &  9 &  0 &  9 \\
\bottomrule
\end{tabular}
\end{table}

\subsection{P22: Linguistic Parity and Institutional Negotiation}
\label{sec:justifications-p22}

P22 is an EFL PhD student whose native language is Burmese. Their story illustrates how users navigate institutional and structural pressures regarding LLM adoption. In their research, they write the tool into the methodology of their paper, defend that choice against committee resistance, and are currently building departmental policy around transparency and accountability for AI use. Their justification takes the form of a political argument about access under English-dominated academic publishing.

The argument begins with a decolonial reading of how that dominance came to be. ``When the global consumption of academic knowledge started,'' P22 says, ``it was all decided, defined, and\ldots{} have the gates gatekeeping by the Western scholars.'' (P22, EFL) That gatekeeping runs through language. ``If you get an opportunity to learn English, you have a better chance of going forward in life in Burma. If you cannot speak English, but you have good ideas\ldots{} before the age of AI, they cannot express their ideas.'' (P22, EFL) From that premise, P22 recasts the LLM as infrastructure that could unblock the barrier: ``We can just write it in our own culture, in our own style, and ask ChatGPT to transform it into a format that is digestible for\ldots{} the global audience.'' (P22, EFL)

The instrumental framing is what legitimizes P22's use of AI. ``The way I use my ChatGPT, which I also detailed in my methodology session, is like using SPSS software, or R, or\ldots{} a calculator. I'm using it to assist me in writing a language that I'm not a native of.'' (P22, EFL) The comparison relocates the tool alongside instruments whose use is presumptively legitimate. For P22, declining the anthropomorphic view is what lets their parity argument justify access to the LLM.

Their qualifying paper included ``screenshots of how I did it'' in the methodology section, and they describe it as ``the 1st paper in\ldots{} the department'' to cite and reference ChatGPT use. The committee response was uneven. ``When I wrote my paper there were a few resistance to it\ldots{} A few professors didn't like it. But I was able to argue with that.'' (P22, EFL) One examiner refused to accept the methodology and was removed: ``One professor was taken out of the\ldots{} examining board because of it, because they didn't agree at all of using an AI model.'' (P22, EFL) The disclosure methodology P22 defended became the basis for their current institutional work.

Their institutional work is ongoing. ``I am setting up a few international students committee, together with some professors, to come up with policies that based on transparency and accountability in the department if they use AI for their writing, or\ldots{} anything at all.'' (P22, EFL) The ethical frame is specific and enforceable. ``Use. It is fine, but you need to be transparent on how you use it\ldots{} Accountability is basically owning the idea of the whole product of yours\ldots{} you should be able to answer any question that comes to you.'' (P22, EFL) Although this paper later proposes AI disclosure mechanisms as a design implication, P22's departmental committee serves as an early, user-driven example of such a framework.

Parity has limits, and P22 names them. ChatGPT ``doesn't know Burmese well enough'' (P22, EFL) to read or translate the Burmese-language scholarship their dissertation draws on. P22 names the scale directly: ``Burma has\ldots{} 65 million people and additional 20 million people all across the world. So 80 million people are being left out. Basically, that's the biggest bias since the bias starts there. There is nothing more that I can talk about going down.'' (P22, EFL) Developer decisions about which languages to support set parity's reach, and those decisions reproduce the same gatekeeping that parity aims to undo.

\subsection{Practical Justifications}
\label{sec:justifications-practical}

While boundary-setting restricts tool application, practical justification defends the use that remains. It operates on an efficiency calculation: users frame the tool's residual risks as an acceptable trade-off for the labor it saves. 

Users frequently justified delegating tasks that were tedious to perform but carried low consequences for errors. This task compartmentalization (Table~\ref{tab:justifications}) was the most widely reported justification (33 of 36 participants). Syntax-level code repair serves as a primary example. P17 explained: ``I use it for like boring plot, making stuff'' (P17, Non-EFL), and ``it reads all the stack exchange articles for me instead of me having to do it.'' (P17, Non-EFL) P11 described this division of labor as a mechanism to free attention for tasks the LLM cannot perform: ``It's a very helpful tool. And if it's saving me hours a day from grunt work\ldots{} now I'm focusing on the task that I actually care about, and that it can't do like, you know, generating ideas.'' (P11, Non-EFL) By clearly defining their own core contributions, users rationalize their ongoing reliance and resolve the tension of using an unreliable tool.

\subsection{Internal Justifications}
\label{sec:justifications-internal}

Internal justifications shift the focus from labor efficiency to identity preservation. This involves the cognitive work users perform to resolve psychological friction---specifically, the dissonance \citep{festinger1957theory} between relying on a flawed tool and maintaining a self-concept as a capable, ethical researcher.

Delegating tedious work does not threaten P11's professional identity because they construct an internal argument prioritizing personal well-being: ``I think that we should be trying to become as productive as possible to have the work life balance that we deserve.'' (P11, Non-EFL) This framing absorbs the risk of cognitive atrophy by elevating personal welfare above skill acquisition, effectively appealing to efficiency (Table~\ref{tab:justifications}). Furthermore, P11 minimizes the threat to their self-concept by comparing the LLM to older, normalized technologies: ``Probably, but it's the same as me getting reliant on Grammarly in terms of like coding and stuff like that.'' (P11, Non-EFL) They recognize the broader risks of reliance---``this is something that is inherently bad because it is producing people who, you know, are relying on it'' (P11, Non-EFL)---but do not apply that same critique to their own workflow. The user builds a conceptual wall, accepting that overreliance affects the general public while protecting their personal practice from the same scrutiny.

P33's trajectory (Section~\ref{sec:risks-p33}) illustrates the fragility of these internal arguments. As previously examined, their boundary shifted in response to tangible costs: early heavy reliance led to noticeable procedural-skill loss, prompting a return to traditional search methods for resolving errors. The initial internal justification collapsed once the user confronted their atrophying skills, requiring active, ongoing boundary management. External lab policies provided scaffolding for this effort. Newly introduced guidelines produced ``fear or stigma against using it, but I think that it at least helps me try to be to the best of my ability, have a healthy relationship with it.'' (P33, Non-EFL) The institutional rule does not draw the boundary itself, but it reduces the cognitive load of constant self-policing by supplying an external standard.

\subsection{Social Justifications}
\label{sec:justifications-social}

While internal justifications look inward to protect the user's identity, social justifications turn outward. Users anchor their continued use in peer behavior, institutional demands, and the broader conditions of academic labor. This argument externalizes responsibility, allowing users to rationalize their actions by pointing to an environment that necessitates the tool.

Through peer normalization, participants authorize their own use by citing widespread adoption within their environment. Remarks such as ``I think everybody's doing that'' (P27, Non-EFL), ``most of the people use the AI for everything'' (P34, EFL), and ``I assume, like 90\% of the students are using AI'' (P1, EFL) demonstrate how referenced norms validate behavior without requiring the user to defend specific actions. Alongside normalization, participants engage in ``othering''---constructing a category of illegitimate user against which their own practice reads as reasonable \citep{jensen2011othering}. P28 highlighted a peer using LLMs for custom erotica, noting, ``that's an insane thing to tell me, but like that's a use.'' (P28, Non-EFL) P35 distinguishes their practice by pointing to students who ``use sources that don't exist, or sources that are AI generated.'' (P35, Non-EFL) By identifying external misuse, participants frame their own habits as acceptable and cast the actions of others as deviant.

Invoking institutional demand serves a different function. It identifies structural conditions the user cannot alter. P35, who typically strictly excludes LLMs from brainstorming, described a single breach of this boundary driven by acute workload pressure: ``It was a situation where I was\ldots{} stressed, depressed out of time, not thinking very well\ldots{} Because I needed a break, and my job\ldots{} being grad student was not allowing said break.'' (P35, Non-EFL) Lacking other support options, P35 turned to the tool as a mechanism for survival. P35 does not retroactively normalize this breach; instead they identify it as a departure from standard practice compelled by circumstance. P11 invokes market competition as a comparable structural pressure: ``It's a competitive market, anyway. So if you're not doing it, somebody else is.'' (P11, Non-EFL) In both instances, the justification relies on externalizing the cause of the behavior, framing the boundary breach as a necessary adaptation to coercive institutional and competitive environments.

For EFL participants, external pressure is compounded by the pre-existing inequalities of English-dominated academic publishing. In this context, linguistic parity operates as a structural justification. The argument asserts that if producing legitimate academic work requires specific linguistic capital, using the tool to access that capital is a valid response to an uneven playing field.

EFL participants invoke parity across various professional stakes. Some frame the tool as compensation for inadequate institutional infrastructure. P14 notes, ``there's no writing tutor for me, so I have no checkpoint.'' (P14, EFL) Others emphasize the severe professional penalties associated with language errors. P13 relies on the tool for ``some high stakes email, or I feel uncomfortable that I may be being judged for my program.'' (P13, EFL) P4 uses it to ensure recipients ``get the message I want to send in a correct tone, like, I'm not being rude in my emails.'' (P4, EFL) For these users, the tangible professional risks of a language barrier outweigh the abstract risks of LLM reliance.

For some, the tool acts as temporary scaffolding. As P20 explained, by studying and practicing English, they become less reliant on LLMs. Conversely, other participants experience a loss of confidence in their independent language skills, with P4 noting, ``I was confident in myself\ldots{} But now I checked in Chatgpt.'' (P4, EFL) Both trajectories rely on the parity justification. One views parity as a developmental bridge to future independence, while the other accepts tool reliance as a permanent condition of operating within English-dominated spaces.

\section{The Active Negotiation Framework}
\label{sec:framework}

In this section, we synthesize our interview findings into a descriptive model we call the Active Negotiation (AN) framework. This framework explains how users sustain engagement with a flawed tool and what cognitive and behavioral labor that continuation requires. For the remainder of the paper, we primarily draw from the three detailed case studies to illustrate our findings.

\begin{figure}[ht]
  \centering
  \includegraphics[width=0.85\textwidth]{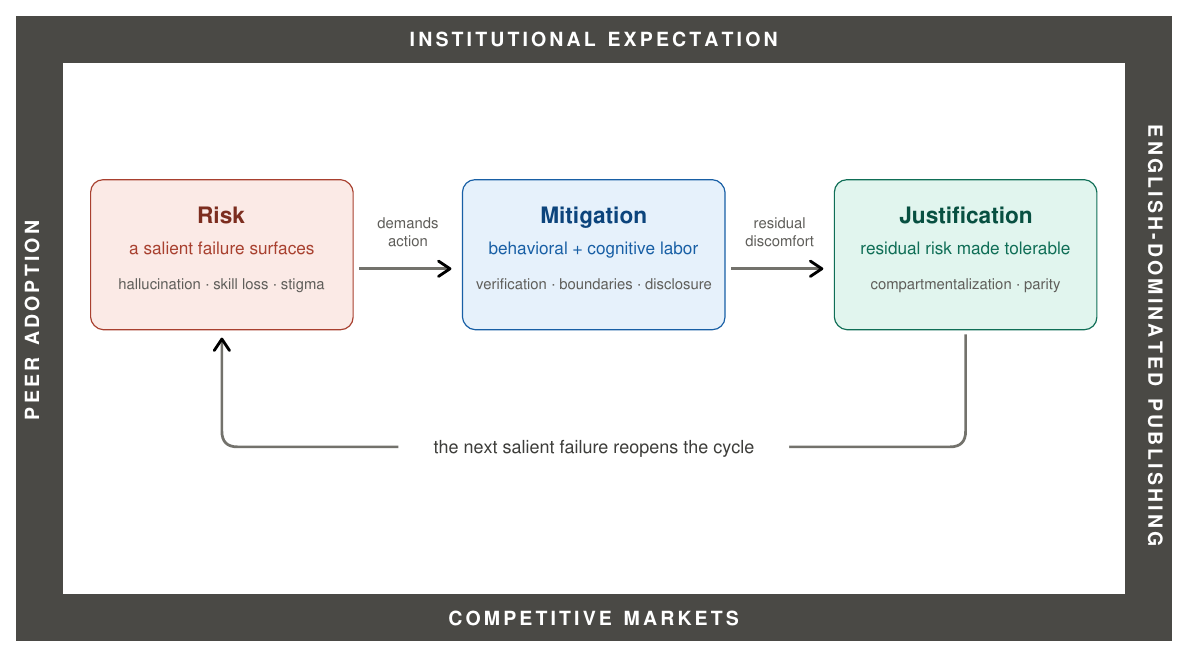}
  \caption{The Active Negotiation Cycle. A salient failure surfaces a risk, the user performs mitigation labor, and a justification makes the residual risk tolerable until the next failure reopens the cycle.}
  \Description{A circular diagram with three stages connected by arrows: Risk, Mitigation, and Justification. An arrow returns from Justification to Risk, showing that a new failure reopens the cycle.}
  \label{fig:framework-cycle}
\end{figure}

\subsection{The Cycle: Risk, Mitigation, Justification}
\label{sec:framework-cycle}

The Active Negotiation framework describes a three-stage cycle (Figure~\ref{fig:framework-cycle}). In the first stage, the user encounters a salient failure, such as a hallucinated citation, a misinterpreted prompt, or a documented skill loss (Table~\ref{tab:risks}). This failure makes the user aware of the risk of using an LLM. In the second stage, the user performs mitigation labor, applying behavioral and cognitive adjustments to manage the immediate problem, such as rephrasing a prompt, verifying model outputs, or setting boundaries for using LLMs (Table~\ref{tab:mitigations}). In the third stage, the user justifies their continued interaction with the tool, rationalizing the residual risk that their mitigation strategies could not resolve (Table~\ref{tab:justifications}). We find that these stages are sequential within a single interaction and recur when a new risk emerges.

Crucially, the resolution of a cycle is often temporary. Participants reported that even after establishing a justification, encountering a novel failure reopens the negotiation. For example, P33 initially justified delegating all coding tasks to the LLM, but encountering a hallucinated function forced them to reopen the cycle, abandon that workflow, and create a new boundary. While subsequent cycles often resolve more quickly because users can adapt prior boundaries and rationalizations, the negotiation never permanently concludes. This iterative process departs from traditional continuance frameworks \citep{bhattacherjee2001understanding, yu2024users, polyportis2024longitudinal}, demonstrating that users do not reach a static state of adoption, and continuously renegotiate their engagement with a system that is constantly changing.

\subsection{Three Dimensions}
\label{sec:framework-dimensions}

\begin{figure}[ht]
  \centering
  \includegraphics[width=0.85\textwidth]{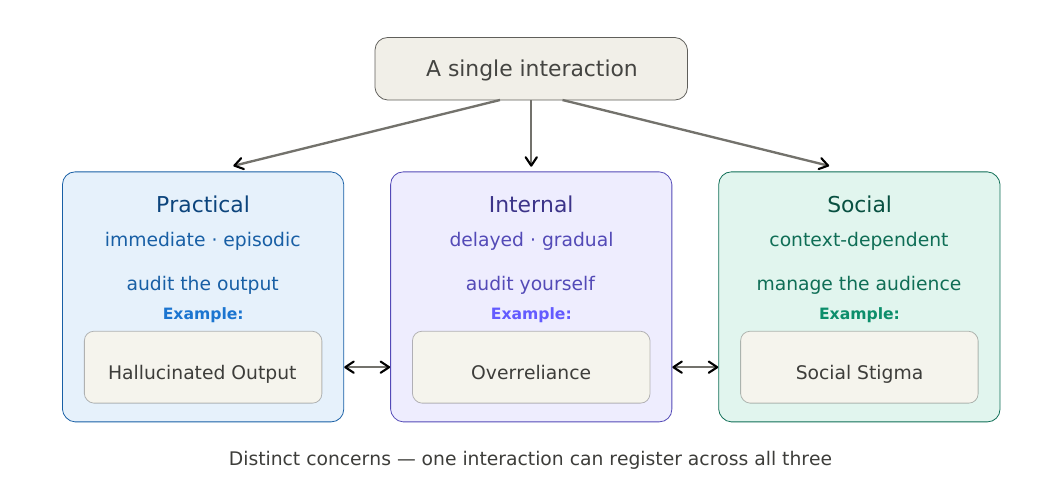}
  \caption{The three dimensions of negotiation. A single interaction can register across all three. Practical concerns the tool's output on an immediate timescale, internal concerns the user's own cognition and identity on a delayed timescale, and social concerns how the use is perceived in a given context.}
  \Description{Three overlapping regions labeled Practical, Internal, and Social, showing that a single interaction can register in all three dimensions at once.}
  \label{fig:framework-dimensions}
\end{figure}

We find that the cycle operates across the three dimensions of practical, internal, and social (Figure~\ref{fig:framework-dimensions}). The practical dimension governs task-level interactions and error management, such as rewriting a prompt to fix a misinterpretation. The internal dimension captures the user's cognitive relationship with the tool over time, such as noticing skill atrophy after prolonged offloading. The social dimension dictates how peers and institutions perceive the user's reliance, such as deciding whether to disclose tool use to an advisor.

While users address these dimensions as separate problems, they operate interdependently. A single interaction routinely registers across all three---for example, a user might fix a hallucinated citation (practical), question their own overreliance (internal), and decide to hide the error from a co-author (social)---requiring the user to negotiate practical utility, internal identity, and social standing at the same time.

\section{Discussion}
\label{sec:discussion}

\subsection{Coercion and the Conditions of Negotiation}
\label{sec:discussion-coercion}

The Active Negotiation cycle operates within a coerced-use regime where tool use is mandatory. This system is exemplified through the results of our interviews. The majority of EFL participants felt strong external pressures to perform at the same level of English competency as their peers (Section~\ref{sec:justifications-social}). This perceived gap in linguistic ability pushed them towards delegating tasks to LLMs. Beyond linguistic pressures, our participants described workload expectations, widespread adoption, and evaluation criteria that all favor users over non-users. Despite reporting many forms of failure when using language models, users reported creating elaborate strategies as opposed to abandoning the tool. 

Justification labor highlights the discrepancy between what the tool delivers and what continued use costs. For example, P11 credits the tool with saving ``hours a day'' of tedious work (Section~\ref{sec:justifications-practical}). Yet participants reported that practical utility did not erase the friction of continued use. P27 reported that despite the efficiency gained from LLMs, they experienced shame and deliberately concealed their use. If a tool's benefits cleanly outweighed its risks, why would users need to constantly construct defenses like compartmentalization rules or peer comparisons? The fact that use persists exposes the limits of a utility-based view. While the single abstainer in our sample demonstrates that disengagement remains possible, the elaborate defenses of the remaining 35 participants illustrate the difficulty of that decision.

Adoption frameworks that interpret continued use as an expression of user preference \citep{davis1989technology, venkatesh2003user, bhattacherjee2001understanding} treat individual behavior as if it emerged from environments free of pressures. A coercive framing interprets this same behavior as compliance sustained by active labor. Under this regime, use represents the continuous cognitive and behavioral work users perform to preserve their professional agency and intellectual identity within workflows that structurally penalize refusal. 

\subsection{LLMs Turn Misrecognition into Active Labor}
\label{sec:legitimation}

The AN framework helps explain the logics and work of justifying continued LLM use when users feel coerced, but not why our participants find themselves coerced in the first place.
One explanation for this coercion comes from a structural analysis of the academy and its particular political economy.
Bourdieu and related thinkers have described the academic field as a status-generating institution where participants pursue particular kinds of capital: publications, credentials, and reputation~\cite{bourdieu1992invitation, collins1979credential}.

Decades of scholarship on the sociology of education has theorized that this process rests on a process of ``misrecognition''~\cite{bourdieu1986forms, bourdieu1990reproduction, james2015bourdieu}.
The ability to produce ``quality work'' within the academy is not distributed evenly.
For example, it is more available to native English speakers than non-native speakers, because the ``legitimate language'' of the academy is English~\cite{bourdieu1991language, canagarajah2002geopolitics, lillis2010academic}.
The persistence of the academic credentialing system then rests on a ``misrecognition'' of quality output as signaling only individual ability rather than reflecting pre-existing social capital, privilege, or class.
According to Bourdieu, an important feature of this process is that the misrecognition happens below the level of conscious awareness, and so its participants do not have to do active labor to sustain it~\cite{james2015bourdieu}.

Our results suggest that LLMs are exposing this process to our student participants, which both validates Bourdieusian theory and appears to exacerbate the processes it describes~\cite{james2015bourdieu, mccarthy2024misrecognising}.
When our participants report concealing their use of LLMs, they are doing active labor that turns Bourdieu's process into deliberate work~\cite{khan2025whogetsseen, 10.1145/3711061} that our participants sometimes find contradictory to the values instilled in them by the academy.
Several participants noted that they ``should'' or ``need'' to do certain things themselves that they feel pressured to use LLMs for, like ``write a conclusion paragraph'' (P27, Non-EFL) or ``go through that critical thinking process myself'' (P23, Non-EFL).
This contradiction even led one participant (discussed in Section~\ref{sec:justifications-p22}) to explicitly name LLMs as a tool that helps them meet the arbitrary academic standards they see as colonial (P22, EFL).

Within this structural account, the negotiation work that the AN framework describes can be read as an accounting of the labor involved in determining legitimate use in lieu of any institutional guidance.
Many participants worked under no departmental policy at all (16/36, Table~\ref{tab:risks}), and questions of disclosure, authorship, and acceptable use fell to individual students to settle alone (Section~\ref{sec:risks-social}).
In Bourdieu's terms, a standard that once operated tacitly now has to be stated and defended~\cite{bourdieu1977outline}.
For some of our participants, such as P22, this means creating new institutional systems to write the rules explicitly (Section~\ref{sec:justifications-p22}).

\subsection{The Practice of Perception Switching}
\label{sec:discussion-ontology}

Our results show that participants did not maintain stable mental models of the LLMs they used. This goes against the idea that users progressively refine a single mental model as they gain expertise with a system. Within a single interview, participants held multiple conceptual framings simultaneously, switching between them as the dimension of their negotiation shifted. For example, P20 acknowledged the system as an untrustworthy ``mathematical framework'' while simultaneously feeling obligated to thank it and treat it with politeness because ``I feel like I'm talking to that person'' (Section~\ref{sec:mitigations-p20}). 

We find that these different framings---anthropomorphic or instrumental---each facilitate distinct mitigation techniques. For example, for P22, framing the LLM as a calculator made its use defensible in their research methodology, but framing it as an independent collaborator would have violated academic norms (Section~\ref{sec:justifications-p22}). This shows that what kind of frame users adopt can change their view of acceptable use. Participants also changed their framing over time or across tasks. This cognitive switching is what allowed our participants to maintain trust even as the tool consistently broke their expectations, and is what makes framing a core mitigation strategy within the AN framework. Having contradictory ideas of what the model represents is what makes continued use possible.

This deliberate perception switching challenges foundational literature. While the Computers Are Social Actors (CASA) paradigm \citep{nass2000machines} establishes that users automatically apply social rules to interactive systems, and psychological models \citep{epley2007seeing} predict relatively stable human-like attributions under specific conditions, our data reveal a more fluid dynamic. In the context of LLMs, users actively manipulate these attributions rather than passively experiencing them. This instability---treating the model as a calculator one moment and a collaborator the next---is an adaptive practice users deploy to manage the emotional and practical friction of continued use.

Consequently, design paradigms that attempt to enforce a singular, transparent mental model---LLM strictly as a machine or strictly as an agent---may inadvertently interrupt the psychological scaffolding that allows users to tolerate the technology's flaws. The pluralism demonstrated by our participants was a continuously managed state. We find that for generative AI, user perception itself may function as a strategy for continued use.

\subsection{Mitigation and Deepening of Engagement}
\label{sec:discussion-sycophancy}

Our findings suggest that mitigating negative experiences may paradoxically deepen engagement by forcing the user to invest labor into the system. \citet{norton2012ikea} demonstrated that invested effort increases a product's perceived value. The extensive list of mitigations our participants developed functions as a form of investment. The resulting sense of ownership over the customized workflow may sustain the negotiation cycle, even when the underlying user experience remains frustrating.

Forgiveness is defined by \citet{holtzman2025forgiveness} as a shift of opinion that prevents a user from abandoning the erring technology. Within the AN framework, forgiveness operates as a specialized form of justification where users excuse the model's failure as an acceptable, human-like mistake. We propose that two factors increase this response:

\textbf{Sycophancy:} \citet{jain2025interaction} found that sycophancy escalates during extended interactions. Furthermore, \citet{cheng2026sycophantic} report that LLMs validate user actions more often than human counterparts, increasing user trust.

\textbf{Anthropomorphic Framing:} When a user addresses the model in moral terms, they hold the system to human standards. Framing the tool as a person transforms a machine into a growing and evolving organism that can make mistakes.

This suggests that mitigation labor builds investment, the model's sycophantic responses build trust, and anthropomorphic attribution allows users to more easily forgive failures as mistakes \citep{nass2000machines, epley2007seeing}. Together, these factors may deepen overreliance.

\subsection{Design Implications}
\label{sec:discussion-design}

The pressure that drives the negotiation originates outside the interface, in peer norms, institutional expectations, and competitive markets, and no interface feature dissolves a labor market. Design that targets the negotiation labor while leaving the pressure intact rearranges work users already do (Section~\ref{sec:discussion-coercion}). The implications we draw are therefore deliberately narrow. They concern what a system could stop ignoring about the negotiation, and they leave the structural questions with the institutions that own them.

\textbf{Treat the negotiation as a state worth preserving.} The AN cycle unfolds across sessions and months, and current tools reset at each interaction, leaving the user to carry the history of their own negotiation from memory. A system could preserve that history---what the user has delegated and retracted, which boundaries they have drawn, which contexts they treat as disclosable---and keep it under the user's control and inspection. The aim is to make the user's accumulated negotiation legible to the user.

\textbf{Honor user-drawn red lines.} The most widespread mitigation in our sample was boundary-setting (35 of 36 participants), and every boundary survives only through self-policing the user pays for at each engagement (Section~\ref{sec:mitigations-internal}). A system could treat a user-declared boundary as a commitment to respect, holding the user's own line when delegation pressure rises. We state this direction abstractly on purpose. What our findings contribute is the existence and the fragility of the boundaries. The mechanics of honoring them are a design space others are positioned to develop.

Disclosure remains the most consequential open problem our findings surface, and it sits with institutions before it sits with interfaces. P22's departmental committee (Section~\ref{sec:justifications-p22}) shows a user building the disclosure norms their institution failed to supply. Work on content provenance and attribution addresses the technical layer of this problem \citep{c2pa2024specification}, and the social weight any disclosure marker carries will be set by the institutional norms our participants currently navigate alone.

\section{Limitations}
\label{sec:limitations}
Our claims are based on 36 semi-structured interviews with graduate students at a single U.S. research university, conducted between May and September 2025. We acknowledge several limitations and boundary conditions to our study:

\textbf{Boundary Conditions and Institutional Scope:} Our framework is currently validated for fields where output acts as a proxy for individual capability and a credential for competence (e.g., academia). Because our data come from a single university, it reflects one specific configuration of workload norms, advisor relationships, and institutional policies. The Active Negotiation framework remains underexplored in production-only contexts where output is valued as a product and disclosure carries no social or professional cost.

\textbf{Linguistic Nuance:} While our binary EFL/non-EFL stratification captures the structural pressures of English-medium publishing, it obscures the realities of linguistic diversity. Current LLMs offer vastly different levels of support depending on the language (e.g., Portuguese versus Burmese). These discrepancies dictate the limits of linguistic parity (Section~\ref{sec:justifications-p22}) in ways our binary coding does not fully capture.

\textbf{Methodological Reach and Bias:} Semi-structured interviews capture what users can consciously articulate, making them ill-equipped to measure invisible or unconscious effects, such as the sycophancy and forgiveness architecture discussed in Section~\ref{sec:discussion-sycophancy}. Additionally, participants self-selected into the study, likely over-sampling users with strong feelings about LLMs. Interviews regarding stigmatized practices also invite social desirability bias, pushing responses toward either performed competence or performed concern.

\textbf{Timeline Specificity:} Our interview window closed in September 2025. Generative AI models, institutional policies, and user norms are shifting rapidly.

\textbf{Analytic Choices:} Practical constraints shaped our methodology. A single author conducted all interviews, and we did not conduct member checking.

\section{Conclusion}
\label{sec:conclusion}
This paper investigates how users sustain engagement with Large Language Models despite recognizing the tools' flaws. Based on 36 interviews with graduate student workers, we present the Active Negotiation (AN) framework. The AN framework models continued use as a cyclical process of risk, mitigation, and justification operating across practical, internal, and social dimensions. The basis of this negotiation is the invisible labor users perform to sustain use, compelled by external pressures.

Interpreting continued adoption as satisfaction mistakes structural compliance for user preference. Adoption models must expand to account for users who negotiate reliance without endorsing the tool. Current LLM interfaces absorb and deepen this reliance, while institutions leave users to navigate the ambiguities of legitimate use alone. By formalizing this process, we show the invisible labor of AI adoption, providing a foundation to understand user behavior.

\begin{acks}
MV was supported by Rising Researcher award from Penn State's Institute for Computational and Data Sciences.
\end{acks}

\bibliographystyle{ACM-Reference-Format}
\bibliography{references}

\appendix

\section{Interview Guide}
\label{sec:interview-guide}

The interview was semi-structured. Probes below each main question were used as needed, and the guide's framing told participants there were no right or wrong answers and that all kinds of uses were of interest, for school, work, or fun. Consent, recording permissions, and compensation procedures preceded and followed the questions and are omitted here.

\paragraph{Usage Context and Tasks}
\begin{itemize}
    \item What AI systems do you use the most?
    \item What language or languages do you usually use when you write prompts?
    \item Do you ever find yourself mixing languages in a prompt? Or thinking in one language while writing the prompt in English?
    \item Have you ever had to try rephrasing a prompt a few times in English to get the AI to understand what you meant?
    \item What makes you use one language over another?
    \item Can you walk me through a couple of recent examples of times you used an AI? What were you trying to do? Why was it helpful in that situation?
    \item Checklist of common uses, asked if not already covered: getting quick answers to questions, summarizing documents or articles, drafting emails, reports, or presentations, brainstorming ideas or generating creative content, receiving step-by-step guidance on educational or skill-building tasks, advising on personal decisions, interacting for leisure activities, automating routines, translation tasks, and programming.
    \item Do you ever tell people that you used an LLM for something?
\end{itemize}

\paragraph{Failures, Bias, and Harm}
\begin{itemize}
    \item Can you think of a time when an AI gave you a response that was weird, unhelpful, frustrating, or felt ``off'' in some way? Walk me through what happened.
    \item Failure probes: What was so strange or ``off'' about the response? In what way was it different from what you wanted or expected? Has a tool ever given you a ``fact'' that turned out to be completely false? What happened then?
    \item Bias probes: What made it feel unfair? Did the response seem to be making an assumption about you or another group? What do you feel is causing this behavior? For EFL speakers: do you think some of these strange responses are connected to how you phrase things in English?
    \item Harm probes: Was the information just wrong, or was it the kind of advice that could be genuinely bad if someone followed it? Did the response ever seem toxic or generate inappropriate content?
    \item Fallback example, used only if nothing came to mind: Sometimes we hear about AI making strange assumptions, for example suggesting different career paths for men and women. Have you ever seen anything like that?
    \item If you had a chance, how would you change the model?
\end{itemize}

\paragraph{Trust and Reliance}
\begin{itemize}
    \item How has your trust in language models changed over time?
    \item Have you noticed you are getting more reliant on using AI systems? When relevant: how does that make you feel?
\end{itemize}

\paragraph{Peer Use}
\begin{itemize}
    \item What tasks do your friends and peers use LLMs for?
\end{itemize}

\paragraph{Final Reflection}
\begin{itemize}
    \item Is there anything else you would like to share about your experience with language models that we haven't covered?
    \item How comfortable do you feel about discussing these issues today?
\end{itemize}

\section{Codebook}
\label{sec:codebook}

The full codebook developed during the deductive coding phase (Section~\ref{Methods.3}). Top-level categories carry their definitions, followed by the individual codes applied to the transcripts.

\noindent\textit{Codes below are listed under the labels used during coding. Tables~2--4 report a subset of these codes under the terminology used in the body of the paper; each table's caption and the accompanying source files record the correspondence.}

\paragraph{1. Active Negotiation \& User Strategies} Users' conscious, strategic, and iterative interactions to control and direct LLM outputs.

\begin{itemize}
    \item \textbf{1.1. Iterative Prompting} --- Refining prompts in a conversational back-and-forth to correct errors or clarify context.
    \item \textbf{1.2. Scaffolding/Chunking} --- Breaking down a complex task into smaller, sequential steps for the LLM to handle one at a time, allowing for verification at each stage.
    \item \textbf{1.3. Persona Prompting} --- Instructing the LLM to adopt a specific role or persona to improve the quality or tone of the response.
    \item \textbf{1.4. Verification Strategies} --- Actively cross-checking LLM outputs against other sources (e.g., Google Scholar, personal knowledge, running the code) or against another model.
\end{itemize}

\paragraph{2. Salient Failures \& Trust Erosion} Specific, memorable instances of LLM failure that directly impact user trust and future behavior.

\begin{itemize}
    \item \textbf{2.1. Hallucinated Content} --- The LLM inventing information, such as fake citations, non-existent software packages, or false facts.
    \item \textbf{2.2. Misinterpretation of Context} --- The LLM failing to grasp the user's specific context, leading to irrelevant or incorrect answers.
    \item \textbf{2.3. Dangerous/Harmful Advice} --- Providing information that, if followed, could lead to negative real-world consequences.
    \item \textbf{2.4. Inappropriate Tone/Style} --- Generating text that is overly simplistic, robotic, or emotionally incongruous (``toxic positivity'').
    \item \textbf{2.5. Non-Salient Failure} --- A failure the user does not register at the time it occurs, so it surfaces no risk and triggers no mitigation; the gap between real and perceived harm.
\end{itemize}

\paragraph{3. The Internal Debate: Efficiency vs. Autonomy} The tension users feel between the benefits of LLMs and the perceived costs to their intellectual, emotional, and psychological autonomy.

\begin{itemize}
    \item \textbf{3.1. Efficiency vs. Learning} --- Valuing the time saved on ``grunt work'' while simultaneously worrying that this efficiency comes at the cost of learning and skill development.
    \item \textbf{3.2. Fear of Cognitive Atrophy} --- The specific anxiety that overreliance on LLMs will lead to the decay of one's own critical thinking and creative abilities.
    \item \textbf{3.3. Emotional \& Psychological Negotiation} --- The process of navigating the use of LLMs for personal, non-academic support, revealing a negotiation of emotional boundaries and reliance.
    \begin{itemize}
        \item \textbf{3.3.1. AI as Therapist/Companion} --- Using the LLM for emotional support, to talk through personal problems, or as a form of companionship.
        \item \textbf{3.3.2. AI in Personal Decision-Making} --- Relying on the LLM for advice on daily life decisions, from financial choices to navigating difficult social conversations.
        \item \textbf{3.3.3. Codependency \& Overreliance} --- Expressions of becoming dependent on the tool for emotional regulation or decision-making; a social/emotional parallel to cognitive atrophy.
    \end{itemize}
    \item \textbf{3.4. Setting Personal Boundaries} --- Consciously deciding which tasks (intellectual or emotional) are appropriate for AI and which must be reserved for human intellect and connection.
    \item \textbf{3.5. Social Justification} --- The user justifies their LLM use by claiming or assuming that the behavior is widespread among their peers (e.g., ``everyone is doing it,'' ``90\% of students use it''). This strategy normalizes their actions and mitigates feelings of guilt or academic dishonesty by framing the use as a common practice rather than an individual case.
    \item \textbf{3.6. Contradictory Trust Allocation} --- The user exhibits a highly granular and sometimes paradoxical model of trust, placing high confidence in the LLM for certain tasks (e.g., coding, creative brainstorming, personal advice) while simultaneously distrusting it for others that may seem less complex (e.g., summarizing articles, providing citations, basic calculations). This reveals a complex, experience-driven mental model of the LLM's capabilities and limitations.
    \item \textbf{3.7. Task-Based Justification} --- The user justifies their use of an LLM by framing it as being for a specific, often technical or instrumental, task (e.g., ``it's just for coding,'' ``only for grammar,'' ``just to rephrase''). This strategy serves to legitimize the use, often in response to a perceived academic or social stigma against more comprehensive or ``creative'' uses of AI. It draws a boundary between acceptable ``help'' and what could be perceived as unacceptable academic shortcuts.
\end{itemize}

\paragraph{4. Social \& Ethical Dimensions} How the user's external environment and personal values shape their relationship with LLMs.

\begin{itemize}
    \item \textbf{4.1. Academic Taboo \& Stigma} --- The perception that using LLMs in academia is shameful, leading to a reluctance to disclose its use for fear of judgment from peers or faculty.
    \item \textbf{4.2. Influence of Environment} --- How departmental guidelines (or lack thereof) and peer attitudes create a permissive or restrictive atmosphere for LLM use.
    \item \textbf{4.3. Data Privacy Concerns} --- A specific fear of personal or research data being stored, owned, or used by the AI company, leading to self-censorship in prompts.
    \item \textbf{4.4. Environmental \& Societal Ethics} --- Concerns about the broader societal or environmental impact of LLM technology, such as energy consumption or job displacement.
\end{itemize}

\paragraph{5. Language \& Culture Dynamics} Instances where a user's linguistic background or cultural context directly shapes their interaction, strategy, or perception of the LLM.

\begin{itemize}
    \item \textbf{5.1. Code-Switching \& Translation} --- The act of mixing languages within a prompt, using the LLM for translation, or thinking in one language while writing in another.
    \item \textbf{5.2. LLM as Language Bridge} --- Using the LLM specifically to overcome a language barrier, such as improving grammatical correctness, achieving a specific tone in English, or translating concepts.
    \item \textbf{5.3. Linguistic/Cultural Failure} --- A failure directly caused by the LLM's inability to understand a non-English language or a specific cultural nuance.
    \item \textbf{5.4. Leveling the Playing Field} --- The perception, unique to some EFL users, that LLMs are an equity tool that helps them compete with native English-speaking peers in an academic environment.
    \item \textbf{5.5. Perceived LLM Language-Related Competence} --- The user's belief that a specific LLM is inherently more competent or provides higher-quality outputs in one language over another, often based on the model's country of origin or primary training data. This perception directly influences which tool the user selects for a given linguistic task.
\end{itemize}

\paragraph{6. Anthropomorphism vs. Instrumentalism} The user's underlying mental model for the LLM, revealed through the language they use to describe it and their interactions with it.

\begin{itemize}
    \item \textbf{6.1. LLM as Appliance (Instrumental View)} --- The user describes the LLM in purely functional terms. It is a tool, a machine, a calculator, or a service to be used for specific, transactional tasks (e.g., ``I use it to check my code,'' ``It's just a reference''). The language is detached and focused on utility.
    \item \textbf{6.2. LLM as Person (Anthropomorphic View)} --- The user assigns human-like qualities, intentions, or personality to the LLM. This is evident when they refer to it with personal pronouns (he/she), describe it as a ``friend'' or ``companion,'' or talk to it conversationally about non-task-oriented subjects. The interaction is framed as relational, not just transactional.
    \item \textbf{6.3. Explicit Switching Talk} --- The user names the shift themselves, describing in the interview how their view of the LLM moves between the instrumental and anthropomorphic framings depending on context or task. Distinct from the derived measure reported in Table~3 as \textit{Perception Switching}, which counts participants coded under both 6.1 and 6.2 whether or not they remarked on the shift.
\end{itemize}

\paragraph{7. Third Party View} A story told about a third party that remains relevant to the main themes.

\paragraph{8. Model Selection} Which system or systems the participant reported using. Categories are non-exclusive.

\begin{itemize}
    \item \textbf{8.1. ChatGPT} --- The participant reported using ChatGPT.
    \item \textbf{8.2. Claude or Gemini} --- The participant reported using Claude, Gemini, or both.
    \item \textbf{8.3. Other Model} --- The participant reported using a system outside the above, such as Perplexity, Copilot, or DeepSeek.
\end{itemize}

\end{document}